\documentclass[preprint,showpacs,preprintnumbers,amssymb]{revtex4}
\usepackage[T1]{fontenc}
\usepackage{amsmath}
\usepackage{amsthm}
\usepackage{graphicx}

\usepackage{color,graphics,graphicx,fancybox,fancyhdr}
\usepackage{latexsym}

\begin{document}

\graphicspath{../Latex} \DeclareGraphicsExtensions{.eps,.ps}

\title{Microwave intermodulation distortion of MgB$_{2}$ thin films }

\author{G. Lamura,\footnote[1]{Present address: I.N.F.M. COHERENTIA and Dipartimento Scienze Fisiche,
Universit\`{a} di Napoli Federico II, I-80125, Napoli, Italy.} A.
J. Purnell, L. F. Cohen\footnote[2]{Corresponding author. E-mail
address: l.cohen@ic.ac.uk}}

\affiliation{Physics Dept., ICSTM, London SW7 0BA, UK.}

\author{A. Andreone, F. Chiarella, E. Di Gennaro, R. Vaglio}

\affiliation{I.N.F.M. COHERENTIA and Dipartimento Scienze Fisiche,
Universit\`{a} di Napoli Federico II, I-80125, Napoli, Italy.}

\author{L. Hao, J. Gallop}

\affiliation{National Physical Laboratory, Queens Rd., Teddington,
TW11 0LW, U.K.}

\date{\today}

\begin{abstract}

The two tone intermodulation arising in MgB$_{2}$ thin films
deposited in-situ by planar magnetron sputtering on sapphire
substrates is studied. Samples are characterised using an
open-ended dielectric puck resonator operating at 8.8 GHz. The
experimental results show that the third order products increase
with the two-tone input power with a slope ranging between 1.5 and
2.3. The behaviour can be understood introducing a mechanism of
vortex penetration in grain boundaries as the most plausible
source of non linearities in these films. This assumption is
confirmed by the analysis of the field dependence of the surface
resistance, that show a linear behaviour at all temperatures under
test.

\end{abstract}



\maketitle

The recent discovery of superconductivity in MgB$_{2}$
\cite{NAMAG} has raised a worldwide interest in the scientific
community because of its promising perspectives. In particular,
applications of superconducting electronics where the use of
closed cycle refrigerators is mandatory (such as hybrid receivers
for telecommunications) could strongly benefit from this medium
T$_{c}$ superconductor. In this respect, the study of harmonic
generation and intermodulation distortion (IMD) is a fundamental
characterization tool for both the performance of passive devices
in the microwave region and the understanding of the non linear
microwave properties of magnesium diboride. Classically, we expect
that a third order product signal arising in a non-linear
responsive material should vary as the third power of the input
signal. In superconducting materials, this has been observed in
low temperature superconductors like niobium and niobium
nitride.\cite{OATESI,MONACO} High temperature superconductors
(HTS) follow the expected behaviour but at very low power levels
only,\cite{ANDREOI,DIETE} whereas at higher power they usually
show a quadratic dependence of IMD on the input
signal.\cite{ANDREOI,DIETE,FINDI,BALAM,OATESII,WILKER} This has
been generally ascribed to vortex penetration in weak
links.\cite{FINDI,BALAM,OATESII}

In this letter, we present a detailed characterization of the
microwave nonlinearity in MgB$_{2}$ performed by single tone and
two tone measurements. In particular, we show that the IMD
products have the same unusual behaviour found in HTS materials.
These experimental results can be explained in the framework of a
granular model of Josephson coupled grains.

We studied three MgB$_{2}$ thin films with inductively measured
critical temperatures ranging between 20 and 31 K. These films
have been realized using a fully in situ two step approach, by a
planar magnetron sputtering technique in a UHV environment.
Briefly, an amorphous Mg-B precursor is grown at room temperature
on r-cut single crystal 10x10 mm$^{2}$ sapphire. After deposition,
using a simple in situ manipulator, the sample is placed in a
niobium box containing small amounts of Mg and then heated up to
800 $^{o}$C. The box is closed using an indium gasket that
guarantees hermetic sealing. The process is conducted in saturated
Mg-vapor as in a ex-situ process, however it gives a surface film
quality and reproducibility that are typical of an in situ
process. Further details are reported elsewhere.\cite{VAGLIO} The
film thickness, measured by Focused Ion Beam (FIB), is 0.5 $\mu$m
$\pm$ 10\%. X-ray diffraction $\theta-2\theta$ measurements
indicate a wide c-axis orientation of the surface.\cite{VAGLIO}
Film morphology is investigated by Atomic Force Microscopy (AFM)
and Scanning Electron Microscopy (SEM) analyses. AFM shows
granular features at the surface, with average roughness of about
20 nm. SEM images with microprobe analysis show high composition
uniformity over the full substrate area. The films are labelled
from $\#$1 to $\#$3 with increasing critical temperature. The main
superconducting properties of all the samples under test are
summarized in table I.

To investigate the power dependence of the microwave properties of
this new superconductor, we performed in the same system
configuration the measurement of the two tone intermodulation
products and of the surface impedance Z$_{S}$=R$_{S}$+jX$_{S}$ as
a function of the input power. We used an open-ended dielectric
single-crystal sapphire puck resonator (with a TE$_{011}$ mode
resonant frequency of 8.8 GHz) in close proximity to the film with
a variable puck-to-sample distance. The experimental configuration
was experimentally chosen observing at room temperature the
distance that produces the largest perturbation in the cavity
resonance. This procedure made us confident that each film has
been positioned where the magnetic field is close to its maximum
value. The enclosure of the resonator is made of OFHC copper and
cooled by a standard two stage Gifford-McMahon cryocooler system
from T$_{C}$ down to 12K. For the measurement of IMD products, two
pure frequencies f$_{1}$ and f$_{2}$ ($>$f$_{1}$) with equal
amplitudes were generated by two phase-locked synthesizers. The
two signals were combined and applied to the resonant cavity. The
frequencies were separated symmetrically about the center
frequency of the resonant cavity by an amount ($\delta f$) such
that both frequencies were well within the 3-dB bandwidth of the
resonator. All the IMD data presented in this paper are taken with
$\delta$f=10 kHz while the resonance bandwidth was generally not
less than 170 kHz. The output signals of the resonator (the two
main tones f$_{1}$ and f$_{2}$ and the two third order IMD's at
2f$_{1}$-f$_{2}$ and 2f$_{2}$-f$_{1}$) were measured as a function
of the input power by using a spectrum analyzer. Note that the
measurement system does not involve an amplifier since the
non-linearity inherent in a high gain active device would be too
high to allow sensitive measurement of the intermodulation
generated in the superconducting film. Further details are given
elsewhere.\cite{HAOtec,HAO,HAOII}

All the samples under test showed a similar behaviour as a
function of power, therefore for the sake of clarity we will show
the experimental results obtained on sample $\#$3 only. In Fig. 1
the output power at the main tones and the IMD third order
products as a function of the circulating power P$_{circ}$ are
presented at different temperatures. In the inset P$_{out}$ vs
P$_{circ}$ for the samples $\#$1 and $\#$2 at t=0.57 and 0.62
respectively is shown for comparison. The power circulating in the
cavity is evaluated using the following
expression:\cite{LANCASTER}
\begin{eqnarray}\label{1}
P_{circ}=10Log\left(\frac{2Q_{L}}{10^{-\frac{IL}{20}}}\right)+
P_{out}
\end{eqnarray}
where Q$_{L}$ is the loaded quality factor of the cavity, IL
represents the insertion losses and P$_{out}$ is the output power
of the main tone. P$_{circ}$ and P$_{out}$ are expressed in dBm.
There are two main reasons to plot the IMD products as a function
of the circulating power instead of the more conventional input
power: \textit{i)} the behavior does not depend either on the
input coupling or on the resonator quality factor, thus making the
comparison amongst different devices simpler; \textit{ii)}
P$_{circ}$ is directly proportional to the square of the mean
amplitude of the microwave applied field $H_{r.f.}$, that is the
relevant quantity for the study of the intrinsic properties in
superconducting samples. $H_{r.f.}$ is expressed by the following
relation:\cite{LANCASTER,HEIN}
\begin{eqnarray}\label{2}
H_{r.f.}=\sqrt{\frac{P_{circ}}{2\Gamma A }},
\end{eqnarray}
where P$_{circ}$ is expressed in Watt, A is the sample surface and
$\Gamma$ the geometrical factor of the cavity, experimentally
determinated by using as a reference a superconducting YBCO sample
of known surface resistance. In the case of the samples under
test, $\Gamma$ was equal to 403, 494 and 626 $\Omega$ for $\#$1,
$\#$2 and $\#$3 respectively with an error of 15\%. The large
difference in the geometrical factors are due to the different
choices of the puck-sample distance.

In Fig. 1 we observe two important features: \textit{i)} the power
law dependence of the IMD products is constant with the
circulating power and shows an exponent ranging between 1.5 and
1.8; \textit{ii)} the output power of the IMD products decreases
with the increasing temperature. In figure 2 the IMD slopes as a
function of the reduced temperature are displayed for the samples
under study. All values are almost temperature independent with an
exponent ranging between 1.5 and 2.3. The error on the IMD slopes
is estimated between 4 and 8 \%.

In the single tone experiment, the quality factor Q and the
resonant frequency f$_{0}$ of the dielectric resonator are
measured as a function of the applied microwave field $H_{r.f.}$
as calculated using eq. 2. We estimate the changes in the surface
resistance and in the surface reactance by using the simple
general relation:\cite{LANCASTER}
\begin{eqnarray}\label{2}
R_{S}+j\Delta X_{S}=\Gamma
\left[\left(\frac{1}{Q_{0}}-\frac{1}{Q_{0}^{back}}\right)-2j\left(\frac{\Delta
f_{0}}{f_{0}}\right)\right]
\end{eqnarray}
where Q$_{0}$ is the unloaded quality factor when the sample is
inserted in the cavity, whereas Q$_{0}^{back}$ is the unloaded
quality factor of the bare cavity. The error on the surface
resistance and the surface reactance is of 15 \%. In figure 3,
$\Delta R_{S}(H_{r.f.})$ at different temperatures is shown. The
surface resistance linearly changes increasing the microwave
surface field $H_{r.f.}$ until a threshold field H$^{*}$ is
attained. H$^{*}$ represents the field value above which non
linear effects appear. For values larger than H$^{*}$ the line
shape of the resonance peak starts to deviate from a pure
Lorentzian curve, therefore the R$_{S}$ values cannot be correctly
estimated. H$^{*}$ is strongly temperature dependent and its value
is significantly lower than the H$_{C1}^{c}$ bulk values
previously reported for this material.\cite{GARRY}

In conventional nonlinear materials, it is straightforward to see
that third order products must increase with the third power of
input main tones.\cite{OATESI,MONACO} In the case of MgB$_{2}$,
IMD slopes range between 1.5 and 2.3, resembling granular YBCO
thin film behavior.\cite{ANDREOI,DIETE,FINDI,BALAM,OATESII,WILKER}
This unconventional behaviour might have different extrinsic
origins: \textit{a)} a hysteretic process dominated by the
creation and the irreversible motion of Josephson vortices in
grain boundaries;\cite{HALBRITI,HALBRITII} or \textit{b)} heating
effects due to the low thermal conductivity of the film substrate.
The latter source of nonlinearity can be ruled out because of the
high thermal conductivity of sapphire and since for each sample
the IMDs remain unchanged when  $\delta$f is varied between 10 to
50 kHz.\cite{HAO} In support of hysteretic Josephson flux
penetration there are several considerations. As recently
reported,\cite{SAMA} granular MgB$_{2}$ is characterized by
amorphous grain boundaries of metallic character with thickness
ranging between 5 and 20 nm. Thus we can model the system by a
network of S-N-S Josephson coupled grains. In addition, IMD slopes
close to 2 and almost temperature independent can be induced by a
linear dependence of the surface impedance on the applied
microwave field.\cite{BALAM} Indeed this is the observed form of
the nonlinear dependence of R$_{S}$ in our films. In a simple
granular model, the IMD amplitude decreases with the increasing
temperature,\cite{HAOII} as we observe. And finally the
dimensionless parameter $r=\Delta(1/Q)/(-2\Delta f/f_{0})$
\cite{HALBRITI,GOLO} is of the order of unity for all samples (see
table I). All these experimental observations point to a model of
Josephson coupled grains where the main extrinsic mechanism for
microwave losses is the generation and the motion of Johsepson
vortices in grain boundaries. The metallic nature of these grain
boundaries might explain the lack of distinctive evidence for weak
link effects in dc magnetisation measurements,\cite{YURAI} where
proximity effect makes the sample a well connected superconducting
domain without strongly affecting the critical current density
even in reduced critical temperature samples. Indeed the MgB$_{2}$
case may resemble that of YBa$_{2}$Cu$_{3}$O$_{7-\delta}$/silver
composites,\cite{JUNG} where the enhanced intergranular current
density was interpreted as being due to improved coupling between
grains by proximity effect in the intergranular silver.

We are grateful to A. Velichko for fruitful discussions and for
making available the reference YBCO sample used for the
calibration of the experimental set up. We thank A. Berenov and Y.
Bugoslavsky for the measurements of the film thickness. This work
was funded by EPSRC GR/R41514 and GR/M67445 and NPL project
9SRP4040. A. A. gratefully acknowledges the support of the
University of Naples "Federico II" during his staying at ICSTM.

\bibliography{lamura}

\begin{table}

\label{table1}

\caption{Main superconducting properties of the samples under
test. The \textit{r} factor is calculated at the reduced
temperature t=T/T$_{C}$=0.57 for sample $\#$1 and at t=0.5 for the
remaining samples.}
 \label{tableI}
\begin{tabular}{|l|c|c|c|}
\hline
\textbf{sample}&{\textbf{$\#$1}}&{\textbf{$\#$2}}&{\textbf{$\#$3}}\\

\hline

\rm $T_{c}$ [K]                 &   21.2 & 24.2 & 31.4 \\

\rm $\Delta T_{c}$ [K]          &   0.6 & 2.1 & 8      \\

\rm R$_{S}(T=12 K)$ [m$\Omega$] & 13    & 8.9 & 6.1    \\

\rm $r=\Delta(1/Q)/(-2\Delta f/f_{0})$& 0.6 & 0.4 & 0.4 \\ \hline

\end{tabular}
\end{table}

\begin{figure}
\centering
\includegraphics[scale=0.9]{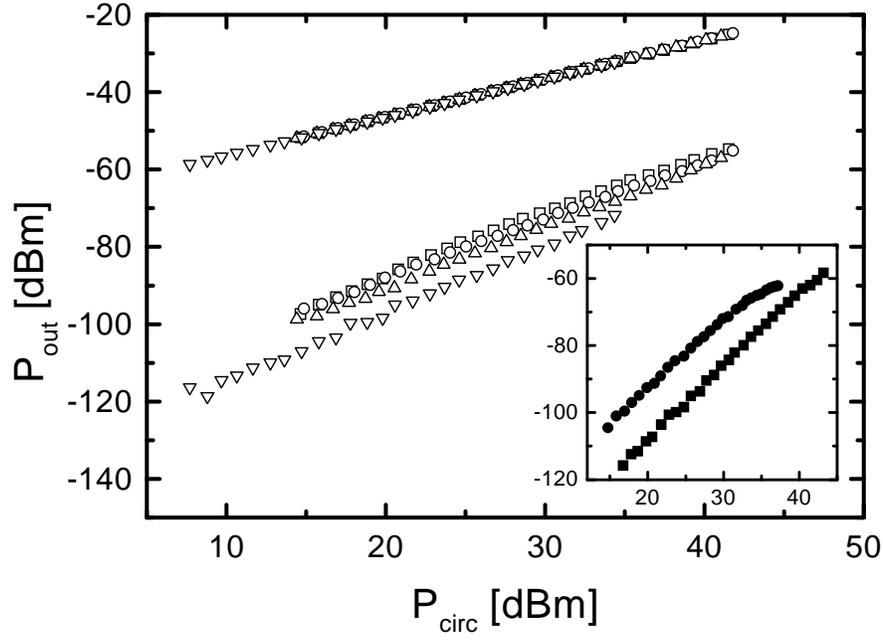}
\caption{The output power of the fundamental signal P$_{out,1}$
and of the IMD products P$_{out,3}$ displayed for sample $\#$3 as
a function of the circulating power at the reduced temperatures:
t=T/T$_{C}$=0.38 (\small{$\square$}\normalsize), t=0.48
(\Large{$\circ$}\normalsize), t=0.57 ($\vartriangle$) and t=0.72
($ \triangledown$). In the inset we show P$_{out}$ vs P$_{circ}$
for the samples $\#$1 and $\#$2 at t=0.57 ($\blacksquare$) and
0.62 ($\bullet$) respectively for comparison.} \label{Figure1}
\end{figure}

\begin{figure}
\centering
\includegraphics[scale=0.9]{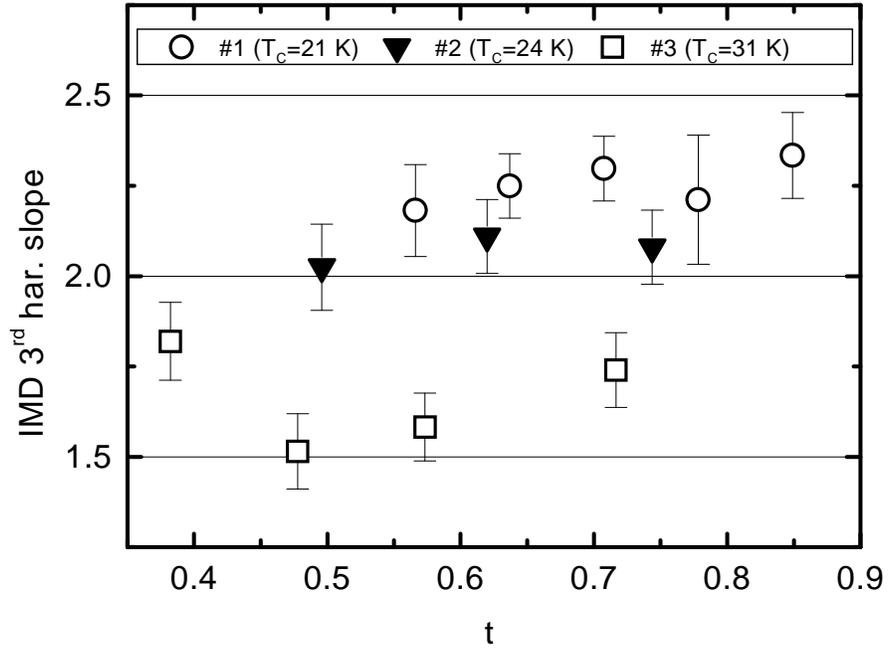}
\caption{The IMD slopes as a function of the reduced temperature
t=T/T$_{C}$ for all the samples under test: $\#$1
(\Large{$\circ$}\normalsize), $\#$2 ($ \blacktriangledown$), $\#$3
(\small{$\square$}\normalsize).} \label{Figure2}
\end{figure}

\begin{figure}
\centering
\includegraphics[scale=0.9]{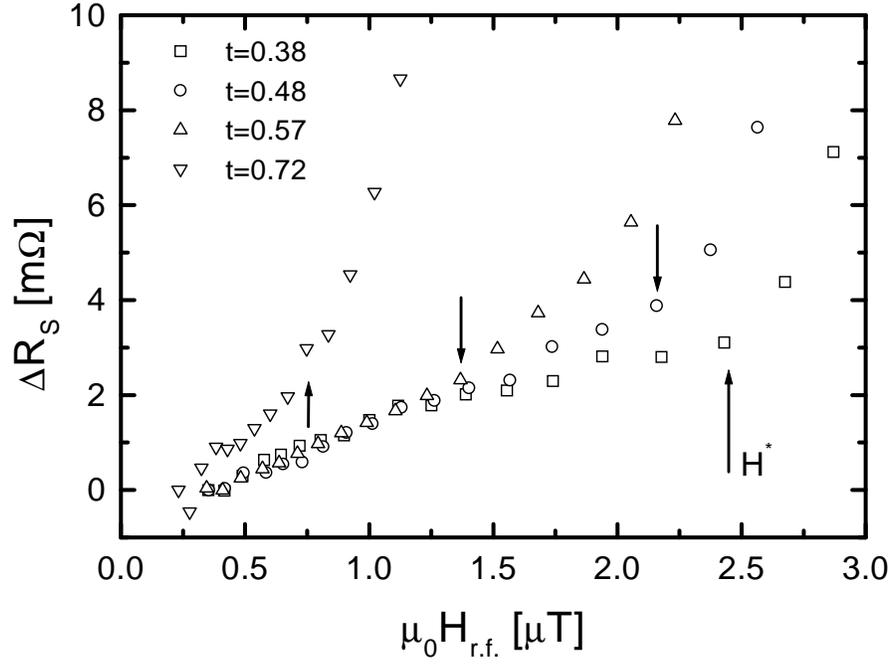}
\caption{Change in the surface resistance values as a function of
the applied microwave field at the following reduced temperatures:
t=0.38 (\small{$\square$}\normalsize), t=0.48
(\Large{$\circ$}\normalsize), t=0.57 ($\vartriangle$) and t=0.72
($ \triangledown$). The arrows indicate the threshold field
$H^{*}$ above which non linear effects starts to show up. }
\label{Figure2}
\end{figure}

\end{document}